\documentstyle[aps,epsf]{revtex}
\begin{document}

\twocolumn [
\hsize\textwidth\columnwidth\hsize\csname@twocolumnfalse\endcsname
\draft

\title{ Quantum Versus Mean Field Behavior of Normal
Modes of a Bose-Einstein Condensate
in a Magnetic Trap}
\author{A. B. Kuklov$^{1,2}$, N. Chencinski$^1$, A. M. Levine$^1$, 
W. M. Schreiber$^1$, and Joseph L. Birman$^2$}
\address{$^1$ Department of Applied Sciences, 
The College of  Staten Island, CUNY,
     Staten Island, NY 10314}
\address{$^2$ Department of Physics, The City College, CUNY, New York,
NY 10031}

\maketitle
   
\begin{abstract}
Quantum evolution of a collective mode of a 
Bose-Einstein condensate containing a finite 
number $N$ of particles shows the phenomena
of collapses and revivals. The characteristic collapse time depends
on  the scattering
length, the initial amplitude of the mode
and $N$. 
The corresponding time values have been derived analytically 
under certain approximation and numerically 
for the parabolic atomic trap. The revival of the mode at time
of several seconds, as a
direct evidence of the effect, can occur,
if the normal component is significantly suppressed.
 We also discuss alternative means to verify
the proposed mechanism. 
\\
   
\end{abstract}
\vskip0.5 cm
]   
   
The progress \cite{ANDERSON,BRADLEY,DAVIS,SCI} in trapping 
the low density alkaline gases and cooling them below the temperature
$T_{BEC}$ of the Bose-Einstein condensation has initiated a search
for features of the atomic coherent collective behavior specific for the trapped condensate.
The theoretical predictions \cite{STRINGARI,EDWARDS,KAGAN} made for the normal
frequencies of the condensate in the traps have been successfully
 confirmed in the experiments \cite{CORNELL,MEWES}
employing the resonant modulations of the trapping potential.

It is worth noting that the calculations of the normal frequencies
\cite{STRINGARI,EDWARDS,KAGAN} are based on the Gross-Pitaevskii (GP)
equation strictly valid in the limit of a number of particles
$N\to \infty$, with the average density $\rho$ being fixed \cite{REV}.
The GP equation describes the condensate in the mean field
(MF) approximation. In this sense the condensate wave function 
$\Phi ({\bf x}, t)$ obeying the time dependent GP equation
is viewed as a classical field \cite{REV}.
 In the atomic traps \cite{ANDERSON,BRADLEY,DAVIS,SCI} neither $N$ can be
obviously considered as infinite nor $\rho $ is fixed ($\rho \sim N$
for $N < 200 - 500$ and $\rho \sim N^{2/5}$ for $N>10^4$ \cite{BAYM}). 
 Therefore, it is important
to understand how the actual quantum evolution of the collective
mode deviates, if any, from that predicted by the mean field approach 
for finite
$N$.
 
In this paper we will show that, while the normal frequencies and
 short time evolution are determined correctly by the GP equation
whenever $N$ is only a few times larger than 1, the quantum behavior
of the normal modes can deviate from the mean field one at times much
longer than the time scale set by the normal frequencies. The quantum
behavior is characterized by dephasing,
that is by the phenomena of the collapses and the revivals
of the normal mode amplitude at zero temperature.
Below we will derive the corresponding quantum solution in the approximation
neglecting the exchange effects. The numerical solution taking into
account the full many body Hamiltonian for $N=300$ in the approximation
allowing $N$ atoms to occupy two single particle levels only is presented also.
It turns out that both approaches yield close values for the collapse
and the revival times, respectively,
 with the collapse time being of the same order
of magnitude as the relaxation time reported experimentally
\cite{CORNELL,MEWES}.

The quantum collapses and revivals were first
analyzed in Ref.\cite{NAROZHNY} for a single atom in a resonant
cavity. Very recently such phenomena have been observed experimentally
\cite{REVIVAL}. As was suggested in Refs.\cite{WALLS,YOU}, the 
phase of the condensate containing finite numbers of
 particles in their ground state should also undergo the
same phenomena.  
 In this case, however, the one-particle density matrix remains
insensitive to such fluctuations. Consequently, special arrangements
of at least two condensates are required in order to detect this effect
\cite{WALLS,WALLS2}.

In contrast, for the condensate 
collective mode the situation is less restrictive. 
The  collapses and revivals of the 
collective mode of the condensate with fixed $N$
 can be detected by the one-particle density matrix 
at zero temperature.  A physical reason for this effect is the following:
For fixed finite $N$ a collective mode is represented by
a multiple mixture 
 of the eigenstates 
rather than by a single eigenstate  of the $N$-body Hamiltonian.
In general, the inter particle interaction makes the eigenenergies dependent
nonlinearly on the quantum numbers of the Fock space. Under such circumstances
the quantum beats develop resulting in collapses and revivals \cite{BEAT}.
For $N\to \infty$, this mixture approaches the coherent state
and can become a true eigenstate of the Hamiltonian. 
Accordingly, in such a limit the beats should vanish.
It is worth noting that
in the atomic traps the existence of this limit is nontrivial because of
the dependence of $\rho$ on $N$. 

Within the Hartree approach  excitations of a Bose-Einstein condensate
can be described by the GP equation
 \cite{REV,STRINGARI,EDWARDS,KAGAN}
$i\dot{\Phi}=\delta E^{(\Phi)}/\delta \Phi^*$ for the classical field
$\Phi ({\bf x},t)$ which represents the Hartree many body
wave function $|\Phi \rangle $ for $N$ bosons. 
Here $E^{(\Phi)}=\langle \Phi|H|\Phi \rangle$
 is the GP
energy functional obtained from the second
quantized Hamiltonian

\begin{equation}
 H=\int d{\bf x} [ \Psi^{\dagger}(H_1-\mu) \Psi +
{u_o \over 2}\Psi^{\dagger}\Psi^{\dagger}\Psi\Psi]
\end{equation}
\noindent
by replacing 
the Bose operator $\Psi$,
obeying the usual Bose commutation rule, by the classical
field $\Phi$ \cite{REV}. As usual, in Eq.(1)  $H_1$ is 
the single particle Hamiltonian; $u_0$ denotes the
interaction constant (we express $u_0=4\pi a/M$ by means of
the positive scattering length $a$ and the atomic
mass $M$ in units where $\hbar=1$); $\mu$ is the chemical
potential. 
In fact, the replacement of the operator $\Psi$ by $\Phi$
is equivalent to projecting of the full  many body wave function
$|\Psi \rangle $ to the Hartree (MF) ansatz \cite{RIPKE}

\begin{equation}
|\Phi \rangle ={(\sum_kC^*_k(t)a^{\dagger}_k)^N|0\rangle 
\over \sqrt{N!}},\, 
\sum_kC^*_k(t)C_k(t)=1,
\end{equation}
\noindent
with $C_k(t)$ being $c$-numbers and
$|0\rangle$ standing for the vacuum state. 
In (2) $a^{\dagger}_k, a_k$ are creation and annihilation
operators, respectively, for a particle at the single particle
state $\psi_k({\bf x})$. 
For the most general representations of the classical field
$\Phi ({\bf x},t) =\sum_k C_k(t)\psi_k({\bf x})$ and
the Heisenberg operator 
$\Psi ({\bf x},t) = \sum_k a_k(t)\psi_k({\bf x})$, 
the replacement procedure
and calculating the mean $\langle \Phi|H|\Phi \rangle $ for the state (2) yield
the same GP functional $E^{(\Phi)}$ from the Hamiltonian (1) 
when $N$ is only a few times larger than 1.

In what follows we will compare predictions for the time evolution
of the normal modes provided by the mean field approach with the
full quantum mechanical calculations which do not rely on the
projection to the states (2).
A good qualitative description of the quantum evolution is provided by 
the following two-level model:
 $N$ bosons can occupy two one particle states, 
one given by the function $\psi_0({\bf x})$ minimizing the
Hartree energy in the ground state
and the other given by the function $\psi_1({\bf x})$ which is orthogonal
to  $\psi_0({\bf x})$. Both functions are taken orthonormal
$\int d{\bf x}\psi^*_i\psi_j=\delta_{ij}$. Keeping $i,j =0,1$ only
restricts the space 
of states available for each particle. 
 In this manner, we neglect the 
processes of interaction
of the collective mode under consideration, which corresponds
to transitions between states $\psi_0({\bf x})$ and $\psi_1({\bf x})$,
with the rest of other modes. Note that the interaction with 
other modes, ascribed to transitions to states
$\psi_k$ with $k>1$, is sensitive to the square of the mode amplitude,
while the rate of the dephasing, as it will be shown, depends
linearly on the amplitude. Therefore, for the small amplitude
the dephasing effect will dominate. 
 
The choice of  $\psi_0({\bf x})$ is dictated by the requirement that
the Hartree variational energy  of the ground state is minimal
for $C_0=1$ (and $C_k=0$ for $k>0$) in (2). 
Taking into account Eqs.(1),(2) and minimizing $E^{(\Phi)}$ with respect
to $\Phi =\psi_0$, one obtains the GP equation  

\begin{equation}
(H_1-\mu)\psi_0 + u_0N|\psi_0|^2\psi_0=0.
\end{equation}   
\noindent
The choice of $\psi_1({\bf x})$, however, depends on the 
characteristic of the collective mode
being studied as well as the magnitude of its frequency \cite{RESTR}.  
In what follows we will employ the Schr\"odinger representation
relying on the time-independent operator $\Psi $. Substituting
$\Psi=a_0\psi_0({\bf x}) + a_1\psi_1({\bf x})$ in
Eq.(1),
one obtains the second quantized Hamiltonian (1) rewritten as 

\begin{equation}\begin{array}{l}
 H=(\varepsilon_{00}-\mu) a^{\dagger}_0a_0  + 
(\varepsilon_{11}-\mu) a^{\dagger}_1a_1 
+[(\varepsilon_{10}+ \\ \\
v(t)) a^{\dagger}_1a_0 + h.c.]
 +{1\over 2}
\sum_{i,j,k,l}u_{ijkl} a^{\dagger}_i a^{\dagger}_j a_k a_l,
\end{array}\end{equation}
\noindent
where $u_{ijkl}= u_0\int d{\bf x}\psi^*_i  \psi^*_j\psi_k \psi_l$ 
($i,j,k,l=0,1$);
 $\varepsilon_{ij}=\int d{\bf x}\psi^*_iH_1 \psi_j$.
We have also introduced
 the term $\sim v(t)$ 
 which models the effect of the modulation of the trapping
potential employed in Refs. 
\cite{CORNELL,MEWES} to excite the condensate normal modes.
In what follows we assume that
$\psi_i$ are chosen in a manner which yields $u_{ijkl}$ real.
 This, in turn, implies the  condition
$u_{ijkl}=u_{klij}$. 

 If the total number of particles $N$ is fixed,
the Hamiltonian (4) becomes
a $N+1$ by $N+1$ Hermitian matrix 
$H_{n,m}=\langle N-n,n|H|m,N-m \rangle$ in the 
Fock space $|n, N-n\rangle $ where $n=0,1...N$ denotes the number 
of atoms at the level $1$, with $N-n$ atoms remaining on the level
$0$. Correspondingly, the Schr\"odinger equation for the many body
wave function in the Fock representation 
$|\Psi\rangle =\sum^N_{n=0}B_n(t)|n,N-n\rangle,\, \sum^N_{n=0}|B_n|^2=1$ is

\begin{equation}
i\dot{B}_n=\sum^N_{m=0} H_{n,m}B_m,
\end{equation}
\noindent
where explicitly the nonzero elements ($H_{n,m}^*=H_{m,n}$) are
$H_{n,n}=\varepsilon n + g_1 n^2/2$,
$H_{n,n-1}=\sqrt{n(N-n+1)}[v(t) +  
(u_{1110}-u_{1000})(n-1)]$, and
$H_{n,n-2}=u_{1100}\sqrt{n(n-1)(N-n+1)(N-n+2)}/2$.
Here the notations $ g_1=  u_{0000} + u_{1111} -4u_{1010}$, 
$\varepsilon = \varepsilon_{11} - \varepsilon_{00}
+ (u_{0000} -u_{1111})/2 +(2 u_{1010} - u_{0000})N$  
are  introduced,
and the overall constant in $H_{n,n}$ is omitted to set $H_{0,0}=0$.
Note that the term $g_1n^2/2$ in $H_{n,n}$ is due to the
interparticle interaction. This describes the effect of renormalizing
of the ideal gas spectrum $\varepsilon n$ in the lowest order
with respect to the scattering length $a$.
 We have also
made use of the relation $\varepsilon_{10} +Nu_{1000}=0$ obtained by means
of  multiplying Eq.(3) by $\psi_1^*$ and integrating over
$\bf x$, given the condition $\int d{\bf x}\psi^*_1\psi_0=0$.
The condensate in this model is associated with the ground eigenstate
of the Hamiltonian matrix in Eq.(5) for $v=0$.
The one particle density matrix $\rho ({\bf x},{\bf x'},t)
= \sum_{i,j}\rho_{ij}(t)\psi^*_i({\bf x})\psi_j({\bf x'})$, with  
$ \rho_{ij}(t)=\langle \Psi|a^{\dagger}_ia_j|\Psi \rangle$
($\rho_{ij}(t)=\rho_{ji}^*(t)$), is given
explicitly by 
 
\begin{equation}\begin{array}{l}
 \rho_{00}(t)=N - \rho_{11}(t),
\, \rho_{11}(t)=\sum^N_{n=0}n|B_n(t)|^2, \\ \\ 
\rho_{10}(t) = \sum^N_{n=0}\sqrt{(n+1)(N-n)}
B_{n+1}(t)^*B_n(t). 
\end{array}\end{equation}

Note that the corresponding MF expression for the density matrix is
 $\rho^{(\Phi)}_{ij}(t)=\langle \Phi|a^{\dagger}_ia_j|\Phi
 \rangle =NC^*_i(t)C_j(t)$,
 where the Hartree ansatz (2) has been employed and
$C_i(t)$ must be found from the  
 GP equation $i\dot{C}_i=\delta E^{(\Phi)}/\delta C^*_i$
in the restricted basis $\psi_0,\,\psi_1$ 

\begin{equation}
i\dot{C}_i=
\sum_j(\varepsilon_{ij}+v_{ij}-\mu \delta_{ij})C_j
+N\sum_{jkl}u_{ijkl}C^*_jC_kC_l.
\end{equation}
\noindent
Here 
$i,j,k,l=0,1$; also we have introduced the notations
$v_{10}= v_{01}^*=v(t)$. This system describes the evolution
of the Hartree state (2) \cite{RIPKE} which is 
expected to be very close to the actual state 
for the case $N\to\infty$
and $\rho$ fixed \cite{REV}.  
For finite and large $N$ one can expect
that the classical and the quantum solutions
do not differ much at short times. Therefore, it is natural
to assume that the driving term 
$v(t)$ in Eq.(7)
 prepares the state (2) (at $t=0$) with some $C_k(0)$ and then is
removed so that the quantum evolution at $t=0$ starts
from the state (2), with $v(t)=0$ for $t>0$. This provides the initial condition
$|\Psi(t=0)\rangle =|\Phi(t=0)\rangle $. Comparing (2) with the full
Fock state $|\Psi \rangle $, one finds 

\begin{equation}
\displaystyle B_n(0)=\sqrt{\frac{N!}{n!(N-n)!}}
\left (C_0^{N-n}(0)C_1^n(0)\right )^*.
\end{equation}

We consider first an approximation when $\rho_{ij}(t)$ can be found
analytically. Specifically, we ignore the off-diagonal
terms of the Hamiltonian (5) which correspond to the exchange of
atoms between the states 0, 1 in the representation (4). Below it will be seen
that this approximation provides qualitatively correct
results when these exchange terms are smaller than the term 
$\varepsilon n$ in $H_{n,n}$ (5). In future work we will show
how this approximation can be extended for the case 
of larger values of the off-diagonal terms. Correspondingly, one has
$B_n(t)=B_n(0){\rm exp}(-iH_{n,n}t)$. 
 This results in Eq.(6)
rewritten as $\rho_{00}=N|C_0(0)|^2=$const, $\rho_{11}=N|C_1(0)|^2=$const
 and

\begin{equation}\begin{array}{ll}
\rho_{10}(t) & =N\rho_1 
\exp (-i(\varepsilon + g_1/ 2)t) \\
&\cdot \left(1+|C_1(0)|^2(\exp (-ig_1t) -1)\right)^{N-1},
\end{array}
\end{equation}
\noindent
with $\rho_1=C^*_1(0)C_0(0)$.
Note that a closed form solution similar to Eq.(9) can be obtained
for a system with arbitrary numbers of levels in the same approximation
neglecting the exchange effects.
We will not
discuss this case here.

For $N>>1$ the short time behavior of Eq.(9) is gaussian, i.e.
$ \rho_{10}(t)=  N\rho_1
{\rm exp}( (-i\epsilon_{01}t-(t/\tau_c)^2)$,
where we have introduced the effective Rabi frequency 
 and the inverse of the collapse time  

\begin{equation}
\epsilon_{01}=\varepsilon_{11} - \varepsilon_{00} +
|C_1(0)|^2Ng_1,\quad
{1\over \tau_c}= {|\rho_1|\over \sqrt{2}}
\sqrt{N}|g_1|,
\end{equation}
\noindent
respectively, and 
$o(1/N)$ terms have been neglected. Note that $\epsilon_{01}$ and $\tau_c$
are sensitive to
the initial conditions given by $\rho_1$. As a matter of fact,
the dependence of $\epsilon_{01}$ on the square of the initial
amplitude ($|C_1(0)|^2\approx \rho_1^2$ for small $|C_1(0)|$) agrees
with the observations of the parabolic dependence
of the mode frequency on the response amplitude 
\cite{CORNELL,MEWES}. Furthermore,
 $\tau_c$ in Eq.(10) is a nonanalytical
function of  $\rho_1$ as well as the interaction
constant $ g_1 \sim u_0$. This implies that the effect
of the dephasing can not be derived by any sort of
perturbation expansion starting from the noninteracting
picture. 
Note also that the classical Eq.(7) in the same
 approximation gives
 $\rho^{(\Phi)}_{00}=\rho_{00},\, 
\rho^{(\Phi)}_{11}=\rho_{11}$ and 
$\rho^{(\Phi)}_{10}(t>0)=N\rho_1{\rm exp}(-i\epsilon_{01}t)$.
Of course, this behavior is purely harmonic 
and demonstrates no signs of dephasing. At short times
$t<< \tau_c$ the MF solution $\rho^{(\Phi)}_{10}(t)$ basically
coincides with the quantum mechanical expression (9). 
 
The solution (9) is suggestive of another peculiar
feature -- the revivals \cite{NAROZHNY,WALLS} of the Rabi 
oscillations with the period
$\tau_r= 2\pi /g_1 >> \tau_c$.
Below we will show that these main features of the analytical
approximation which neglects the exchange terms remain qualitatively
correct for the exact numerical solutions of the quantum as well as
classical Eqs.(5), (7), respectively.

We have diagonalized the Hamiltonian (5) numerically and then calculated
the density matrix (6) for the initial condition
(8). For $N<200$ it is possible to do this directly. For larger
$N$ the required calculation time grows significantly
with $N$. Utilizing the fact that the initial condition
(8) peaks around some value of $n$ allowed us to reduce
the size of the matrix (5). As a matter of fact, the
distribution (8) is the square root of the binomial distribution
whose width is $|C_1|\sqrt{1-|C_0|^2}\sqrt{N}$. This implies that
for large $N$ and small $C_1$ the effective size of the
 Hamiltonian matrix grows
as $\approx |C_1|\sqrt{N}$ instead of as $N$.
Accordingly, we reduced the size of the matrix (5) around the peak
values of the distribution (8) and  ran the program for several
successively increasing sizes until the final solution 
did not change.  The result is shown in Fig.1 for
two different initial amplitudes. For comparison,
we have also plotted the envelope of the classical solution of Eq.(7)
found numerically as well.  
The Hartree ground state function $\psi_0$ and
the function $\psi_1$ are represented
by the variational ansatz $\psi_{000}({\bf x})=
(M/\pi)^{3/4}\omega_z^{1/4}\omega_{\perp}^{1/2}
{\rm exp}(-M\omega_{\perp}(x^2+y^2)/2
-M\omega_zz^2/2)$ \cite{BAYM} 
and the state
  $\psi_{110}({\bf x})=(M\omega_{\perp}(x^2+y^2)-1)\psi_{000}({\bf x})$,
respectively. Here $\omega_z,\,
\omega_{\perp}$ stand for
the variational frequencies \cite{BAYM} along and perpendicular to the
axial axis $z$, respectively. These states were taken for given $N$.
 Accordingly, one finds for $g_1$ in Eqs.(5),(9), (10)
$g_1=-a \sqrt{M\omega_z}\omega_{\perp}/ \sqrt{2\pi}$.
The bare energy of the
single particle transition is $\varepsilon_{11}-\varepsilon_{00}
=\omega_{\perp}+\omega_0^2/\omega_{\perp}$. The solutions for
$\omega_z,\,\omega_{\perp}$ 
were taken from Ref.\cite{BAYM}, with
the frequency $\omega_{\perp}$ at $N=0$ being
$\omega_0=2\pi \cdot 132$Hz
\cite{CORNELL}.
When $N<300$ such a choice corresponds to the lowest breathing mode,
which for large $N$ should transform into that found in Ref.\cite{STRINGARI}.
For $N>500$, the use of the pair of states
$\psi_{000},\,\psi_{110}$
for calculating the matrix elements in (4) becomes a poor
approximation for the actual breathing mode \cite{STRINGARI}.
Therefore in this paper we will limit our analysis
to the case $N=300$. 

We calculated the quantum mean square radius 
$R^2 (t)=(N r^2_{00})^{-1}\sum_{i,j}r^2_{ij}\rho_{ij}(t)$
 in the $xy$-plane
of the condensate 
in units of $r^2_{00}=1/M\omega_{\perp}$, where explicitly
$r^2_{00}=\int d{\bf x}\psi^*_{000} (x^2+y^2)\psi_{000}$,
$r^2_{11}=\int d{\bf x}\psi^*_{110} (x^2+y^2)\psi_{110}=3r^2_{00}$
and $r^2_{10}=\int d{\bf x}\psi^*_{110} (x^2+y^2)\psi_{000}=r^2_{00}$,
as a physical quantity responding to the
Rabi oscillations.
The MF approach gives $R^2(t)=1+2(|C_1|^2+{\rm Re}(C^*_1C_0))$.
Obviously, in the Hartree ground state ($C_1=0$) this quantity equals 1.
Furthermore, exact calculations of $R^2$ for the true many body ground
state of the Hamiltonian (5) for $N<500$ give a value which is 
very close to 1 as well. For the initial condition (8), one finds that
$R^2(0) -1=2\rho_1+2|C_1(0)|^2$ for both the Hartree state and the actual state 
governed by the Schr\"odinger equation (5). We will analyze
the case of small disturbances of the ground state. This
corresponds to $|C_1|<1/3$ which allows one to take
$R^2(0) -1\approx 2\rho_1$ with a very good precision.
Therefore the quantity
$2\rho_1$ in Eqs.(9),(10) plays the role of the initial amplitude
of $R^2(t)$ for  the small amplitudes. 
\begin{figure}[hbt]
\epsfxsize=\columnwidth\epsfbox{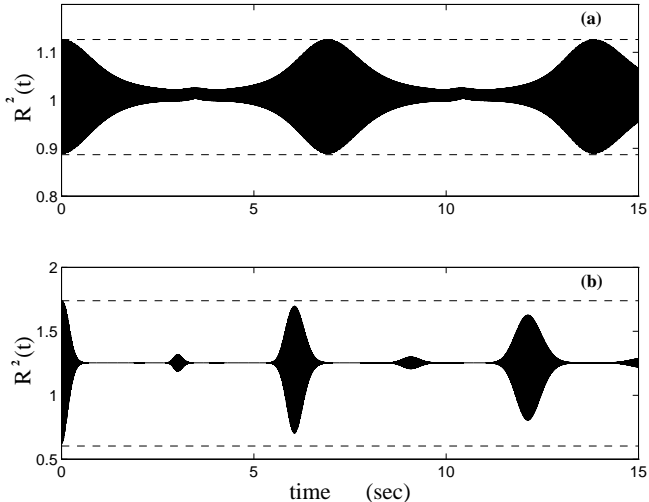}
\caption{ Collapses and revivals of oscillations of the mean square radius $R^2(
t)$ of
the condensate for $N=300$: a) the initial amplitude $2\rho_1= 0.1$;
b) the initial amplitude $2\rho_1= 0.6$. Because of large time scale chosen,
the rapid
oscillations are seen as a black background. The dashed lines correspond
to the envelope of the classical solutions}
\label{fig1}
\end{figure}
Note that the amplitude of the classical
solution (see the dashed line in Fig.1) stays constant and
equal to $2\rho_1$. In contrast, 
 the
collapses and the revivals of the amplitude of the 
oscillations of $R^2(t)$ determined quantum mechanically are clearly
seen in qualitative accord with
 the analytical solution (9).
We have approximated the initial collapse stage of the numerical
solution by
a gaussian fit to the envelope and found the characteristic
collapse time $\tau_c$ 
as a function of the initial amplitude $2\rho_1$.
 The result for
$1/\tau_c$ is represented in Fig.2. 
As can be seen,
this graph is very close to the straight line predicted by Eq.(10).
Its slope is $7.0 s^{-1}$ while the analytical formula
(10) gives $10 s^{-1}$.
These values yield the shortest
dephasing time (corresponding to $\rho_1=1/2$ in Eqs.(9),(10)) as
$\approx 0.1 s$ which is comparable to the relaxation times
($\approx 150$ ms \cite{CORNELL} and $\approx 250$ ms \cite{MEWES}) reported
experimentally.

A direct confirmation of the proposed mechanism would be the 
experimental observation of the revivals. In this regard we
note that the
expected values of
$\tau_r$ are of the order of seconds as given by the
analytical estimates and the numerical solution
(see Fig.1). This is comparable with 
the lifetime of the condensate in the 
traps \cite{ANDERSON,BRADLEY,DAVIS,SCI}.
Furthermore, the normal component which introduces 
irreversible dissipation would inhibit the revivals unless
its concentration is strongly suppressed. We will discuss
this point in greater detail in future work. An estimate 
of the degree of the required suppression can be obtained from
the following considerations. 
The rate $1/\tau_n$ of the thermal dissipation
should be  proportional to the relative amount $p$  of the normal 
component in the trap. In other words $1/\tau_n=p/\tau_0$,
where $\tau_0$ is the relaxation time  
above $T_{BEC}$ when all particles constitute
the normal component ($p=1$). This time is about
$\tau_0\approx 20 $ms \cite{CORNELL,MEWES}.
Consequently, in order to observe the revival of the oscillations
at times $t>\tau_r$, the temperature should be reduced so that
the condition $\tau_n > >\tau_r$ or 
$p<< \tau_0 /\tau_r \sim 10^{-3}$ holds. 

\begin{figure}[hbt]
\epsfxsize=\columnwidth\epsfbox{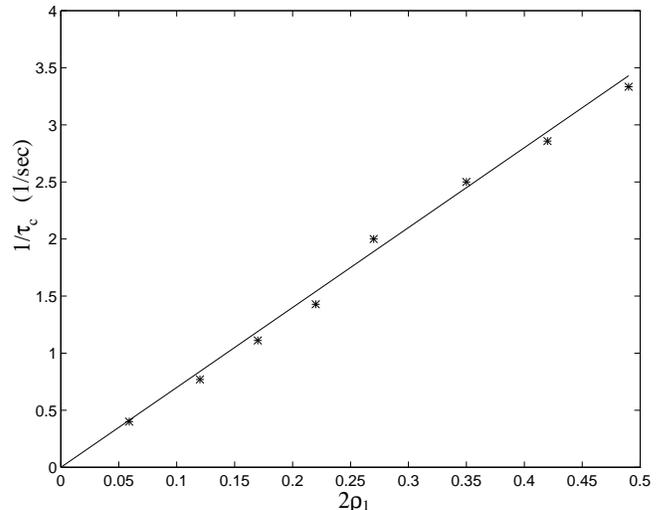}
\caption{ The rate of the first collapse as a function of
 the initial amplitude $2\rho_1$ for $N=300$. The line is a guide for eyes.  }
\label{fig2}
\end{figure}

An indirect indication would be the observation
of the relaxation of the oscillations described by the gaussian
dependence, with the relaxation rate proportional to the initial
 response amplitude (for the small amplitudes) 
as predicted by Eq.(10) and Fig.2.
 Furthermore, Eq.(10) predicts a specific 
dependence on $N$. Let us focus on this fact in detail.
The coefficient $g_1$ in Eq.(10) is given by the effective
volume $V$ occupied by the condensate as $g_1\sim a/V
\sim a\rho /N$.
For small $N$ the density $\rho \sim N$ 
yielding  
$1/\tau_c\sim \sqrt{N}$ in Eq.(10). 
For $N>200 - 500$ the volume $V$ expands 
as a function of $N$ due to the inter particle interaction 
\cite{BAYM}. Accordingly, the
dependence of $1/\tau_c$ should deviate from $\sim \sqrt{N}$.
If $\rho$ were independent on $N$, one would have
obtained $1/\tau_c \sim \rho /\sqrt{N}\to 0$ which
implies no dephasing for large $N$.  In the approximation we employed 
contributions of the higher order terms in $\rho$ to $g_1$
were neglected. Therefore, even though $\rho$ grows
with $N$ ($\rho \sim N^{2/5}$
\cite{BAYM} for large $N$), the dephasing rate, as given by Eq.(10), is expected
to vanish as a slow function of $N$ 
($1/\tau_c \sim N^{-1/10}$) for large $N$. 
The higher order terms in $\rho$ contributing
to the dephasing rate could change this conclusion.
In future work we will analyze the limit of large
$N$ by means of finding a better choice for the
single particle state $\psi_1$ for large $N$ using the Bogolubov
approach.

In conclusion, we have proposed a mechanism of 
 interaction induced dephasing  
of the collective excitations in the atomic traps. 
The calculated collapse time
is comparable to the observed relaxation times for the collective modes.
 Two main features --  gaussian 
relaxation with the time constant depending linearly
on the initial amplitude of the oscillations and
 depending on the total number of atoms 
-- if observed
experimentally, could verify this mechanism. If the temperature
could be reduced much below $T_{BEC}$, the revivals of the collective
oscillations could be seen on times of several seconds. 

We are grateful to Yuri Kagan for very useful discussions.
This research was supported by grants from The City University
of New York PSC-CUNY Research Award Program.


\begin{references}
\bibitem{RESTR}
Furthermore, a frequency of the mode should not
be restricted by the symmetry requirements which
rely essentially on the infinite number of the one particle
levels in the trapping potential as is the case for the
center of mass motion
and for the breathing modes in 2D \cite{PIT}
\end{references}
\end{document}